# Relation between the charge/discharge processes of dust particles and the dynamics of dust clouds over the Moon surface

E.V. Rosenfeld *), A.V. Zakharov **)

*) *Institute of Metal Physics, Ekaterinburg,* evrosenfeld@gmail.com
**) *Space Research Institute, Moscow*

It is shown that the appearance of lunar horizon glow and streamers observed above the lunar terminator described by the dynamic fountain mode byl Stubbs et al. (2006) requires a value of dust particles charge several orders exceed what they obtain on the Lunar surface. To obtain a sufficient charge due to a departure of photoelectrons separated submicron particles have to flow over the surface during tens of seconds or even several minutes. Therefore for emergence of dust streamers at the lunar sunrise it is necessary that dust particles do not lose the positive charge during a night.

**physics.space-ph - Space Physics**

### 1. Введение

По сути дела, электростатическая модель является в настоящее время единственной подробно исследованной теорией, способной объяснить возникновение пылевых облаков над поверхностью астероидов и безатмосферных планет. В этой модели действующая на пылинки подъемная сила имеет кулоновскую природу: пылинка и поверхность планеты заряжены одноименно, так что между ними действуют силы отталкивания. При этом считается, что на освещенной Солнцем части поверхности планеты возникают положительные заряды, поскольку основной ток создается фотоэлектронами. На ночной стороне основной ток создают электроны солнечного ветра, так что заряды оказываются отрицательными [Criswell, Criswell & Do?].

Более углубленное рассмотрение [Stubbs 2006] показывает, что электрическое поле в основном сосредоточено в пределах двойного электрического слоя над заряженной поверхностью планеты. Он возникает потому, что в электрон-ионной плазме по мере приближения к заряженной поверхности растет концентрация частиц с зарядами противоположного знака (экранирование Дебая-Хюккеля). Толщина этого слоя, которая по порядку величины равна Debye length $\lambda_D$, растет с уменьшением плотности плазмы и с ростом ее эффективной температуры.

Поле внутри двойного слоя имеет напряженность $E \approx V/\lambda_D$, где $V$ – разность потенциалов внутри слоя. Оно ускоряет вылетающие с поверхности заряженные пылинки с зарядом $q_d$, одноименным с зарядом поверхности. Если масса пылинки $m_d$, то пройдя двойной слой, она приобретает скорость $v_0$ и кинетическую энергию $\varepsilon_k$, достаточные, чтобы подняться над его поверхностью на высоту $h$:

$$h = \frac{q_d V}{m_d g}, \quad v_0 = \sqrt{2V \frac{q_d}{m}}, \quad \varepsilon_k = \frac{m_d v_0^2}{2}. \tag{1}$$

Здесь $g$ – ускорение свободного падения, и внутри слоя сила тяжести, действующая на пылинку, считается пренебрежимо малой по сравнению с кулоновской силой. Таким образом, если масса пылинки достаточно мала, а заряд достаточно велик, она вполне может подняться на высоту в десятки километров, на которой наблюдается horizon glow.

С принципиальной точки зрения эта теория, предложенная в работе [Stubbs 2006], безупречна, но при попытке подставить в нее конкретные цифры возникают некоторые проблемы, связанные прежде всего с определением заряда пылинки $q$. Действительно, вслед за [?] авторы работы [Stubbs 2006] полагают, что "the charge on a dust grain $q$ is simply given by

$$q = C\varphi_S \tag{2}$$

where $C$ is the grain capacitance", а $\varphi_S$ - потенциал поверхности. Однако, понятие емкости имеет смысл только для металлических тел, да и в этом случае формула (2) верна только если речь идет об *уединенной* пылинке. Если же маленькая пылинка лежит на заряженной поверхности, то равными будут не потенциалы поверхности и (уединенной) пылинки, а поверхностные плотности заряда $\sigma_S$ на обоих телах. Следовательно, мы должны в этом случае написать

$$q = \pi r_d^2 \sigma_S, \tag{3}$$

где $r_d$ – радиус пылинки. Как будет ясно из дальнейшего изложения, заряды субмикронных пылинок в этом случае оказываются на много порядков меньше значений, предсказываемых формулой (2).

Подчеркнем, что цель этого замечания вовсе не в том, чтобы указать на арифметическую ошибку и на невозможность использования понятия "capacitance" в отношении диэлектриков. Дело в том, что вопросы о величине заряда пылинки и, что даже более существенно, о продолжительности периодов времени, в течение которых она заряжается и разряжается, являются важнейшими для понимания динамики пыли. Настоящая работа посвящена обсуждению именно этих вопросов и тех изменений, которые следует внести в существующие представления о динамике пыли, если полученные здесь результаты будут подтверждены дальнейшими исследованиями. При этом мы используем значение $j_{ph} = 4 \cdot 10^{-5}$ A m$^{-2}$ [?] for photoelectron current density, что позволяет максимально упростить задачу. В этом случае фотоэлектронный ток на дневной стороне минимум на порядок превышает токи, создаваемые электронами и ионами солнечного ветра, движущегося со скоростью $V_{sw} \approx 4 \cdot 10^5$ m s$^{-1}$ и имеющего плотность частиц $n \approx 5 \cdot 10^6$ m$^{-3}$ [?]:

$$|j_{sw}| = n|e|V_{sw} \approx 3 \cdot 10^{-7} \text{ A m}^{-2}, \tag{4}$$

Двойной слой над дневной поверхностью планеты в этом случае также создается главным образом фотоэлектронами. Поэтому, пренебрегая в простейшем приближении плазменными эффектами, мы получаем возможность получить многие результаты в аналитическом виде.

Во втором разделе мы подробно рассматриваем процессы зарядки и оцениваем величины зарядов пылинок и плотность заряда на поверхности Луны. В разделе 3 доказывается невозможность возникновения за счет фотоэффекта сколь-нибудь значительного заряда у пылинки, лежащей на лунной поверхности, а в разделе 4 обсуждаются другие механизмы возникновения электрических токов, приводящих к перераспределению зарядов, и возможность сохранения положительного заряда пылинок на продолжении сотен часов в течение лунной ночи.

## 2. Двойной слой вокруг уединенной пылинки

Заряд $q_d$ уединенной пылинки радиуса $r_d$, находящейся в световом потоке, будет расти до тех пор, пока потенциал на ее поверхности[1]

$$\varphi = \frac{1}{4\pi\varepsilon_0}\frac{q_d}{r_d} \approx 10^{10}\frac{q_d}{r_d} \tag{5}$$

не достигнет значения

$$\varphi_0 = K/|e|, \quad K = \frac{1}{2}m_e v_{pe}^2. \tag{6}$$

---

[1] Использование этой формулы предполагает, что плотность вылетающих фотоэлектронов в окружающем пространстве пренебрежимо мала, и что свет падает со всех сторон равномерно. В действительности оба требования не выполняются, но при грубых оценках эти упрощения не играют большой роли.

Здесь $m_e$ – масса электрона, а $v_{pe}$ – начальная скорость вылета, которую мы будем считать одинаковой для всех фотоэлектронов.

После того как потенциал пылинки достигнет значения (6), а ее заряд - величины

$$q_d^{(0)} = 4\pi\varepsilon_0 r_d \varphi_0, \qquad (7)$$

все фотоэлектроны, выбитые с поверхности пылинки, будут тормозиться ее электрическим полем и возвращаться обратно. В результате над поверхностью пылинки возникнет и начнет уплотняться «шуба» (sheath) из улетающих и возвращающихся обратно электронов. Эта «шуба» и положительно заряженная поверхность пылинки образуют двойной электрический слой. В процессе его уплотнения заряд $q_d$ на поверхности самой пылинки продолжает расти, но сумма $q_d$ и заряда $q_e < 0$ окружающего пылинку электронного облака остается неизменной:

$$q_d + q_e = q_d^{(0)}. \qquad (8)$$

Равновесное значение заряда электронной «шубы», которое достигается после ее окончательного формирования, тем больше, чем большее время $T$ каждый электрон тратит на подъем и возвращение на поверхность пылинки. Если обозначить через $j_{pe}$ число фотоэлектронов, которое свет выбивает с единицы площади поверхности за единицу времени, то в равновесии в любой момент $t$ «шубу» будут составлять те электроны, которые вылетели после момента $t$-$T$:

$$q_e = -4\pi r_d^2 j_{pe} T. \qquad (9)$$

Чтобы оценить величину $T$ предположим, что скорости вылета фотоэлектронов $v_{ph}$ не просто одинаковы по величине, но всегда направлены только вдоль радиусов пылинки. Тогда на некотором расстоянии $r'$ от центра пылинки их скорость обращается в нуль, так что полный заряд внутри сферы радиуса $r'$ равен $q_d^{(0)}$ (8). Если считать, что потенциал, как обычно, обращается в ноль на бесконечности, то вне этой сферы мы получаем

$$\varphi(r)\big|_{r \geq r'} = \frac{r_d}{r}\varphi_0. \qquad (10)$$

Внутри этой сферы связь между потенциалом и скоростью $v(h)$ фотоэлектронов на высоте $h$ над поверхностью пылинки имеет вид

$$v(h) = \sqrt{v_{pe}^2 - \frac{2}{m_e}|e|\big[\varphi(r_d) - \varphi(r_d + h)\big]}, \quad h \equiv r - r_d \leq H \equiv r' - r_d, \qquad (11)$$

где $m_e$ - масса электрона. Полное время, которое электрон на пути вверх-вниз проводит в некотором слое толщиной $dh$, равно $2dh/v(h)$, и этому времени пропорциональна плотность вероятности $w(h)$ найти электрон (а потому и плотность заряда $\rho(h)$) на высоте $h$:

$$w(h) = \frac{2}{v(h)T}, \quad \rho(h) = \frac{q_e w(h)}{4\pi(r_d + h)^2}. \tag{12}$$

Для определения потенциала поля мы, как обычно, используем уравнение Лапласа, которое в этом сферически симметричном случае принимает вид:

$$\left(\frac{d^2}{dh^2} + \frac{2}{r_d + h}\frac{d}{dh}\right)\varphi(r_d + h) = -\frac{\rho(h)}{\varepsilon_0}; \quad \varphi(r_d) - \varphi(r') = V \equiv \frac{1}{2}m_e v_{ph}^2. \tag{13}$$

Значение $\varphi(r')$ определяется уравнением (10).

Систему самосогласованных уравнений (9)÷(13) вряд ли можно решить в аналитическом виде. Однако, как легко видеть из (12), плотность электронного заряда резко нарастает когда высота $h$ приближается к $H$, (скорость фотоэлектронов падает) и даже расходится при $h=H$. Поэтому в простейшем приближении можно предположить, что весь заряд электронной «шубы» сосредоточен в тонком слое на самой ее поверхности. Тогда мы получим сферический конденсатор с разностью потенциалов $V=\varphi_0$ и зарядом внутренней обкладки (поверхности пылинки) равным $q_d$, а наружной $q_e$,. Поэтому в слое $0 < h < H$

$$\varphi(h) = \left(\frac{1 - h/H}{1 + h/r_d} + \frac{1}{1 + H/r_d}\right)V, \quad 0 \leq h \leq H;$$

$$v(h) = v_{ph}\sqrt{\frac{r_d}{H}\frac{H - h}{r_d + h}}, \quad H = \frac{q_d^{(0)}}{|q_e|}r_d = \frac{\varepsilon_0 V}{j_{ph}T}. \tag{14}$$

Теперь время полета фотоэлектрона определяется из первого уравнения (12)

$$T = 2\frac{H}{v_{ph}}\left\{1 + \frac{H + r_d}{\sqrt{Hr_d}}\operatorname{asin}\sqrt{\frac{H}{H + r_d}}\right\}, \tag{15}$$

и последняя из формул (14) дает замкнутое уравнение для определения толщины двойного слоя:

$$\frac{\varepsilon_0 V v_{ph}}{2 j_{ph} r_d^2} = x^2\left\{1 + \frac{1 + x}{\sqrt{x}}\operatorname{asin}\sqrt{\frac{x}{1 + x}}\right\}, \quad x = H/r_d. \tag{16}$$

При выбранном нами значении $j_{ph} = 4\cdot 10^{-5}$ A m$^{-2}$ левая часть этого уравнения имеет величину порядка $0.07 V^{3/2}/r_d^2$, где потенциал двойного слоя $V$ составляет несколько вольт. Поэтому больше единицы. Поэтому для толщины двойного слоя в зависимости от радиуса объекта мы получаем очень простые выражения:

$$H \approx \begin{cases} 0.25\cdot V^{3/4} \text{ m}, & r_d \gg 1\text{ m} \\ 0.3\cdot V^{3/5} r_d^{1/5} \text{ m}, & r_d \leq 1\,\mu\text{m} \end{cases}. \qquad (17)$$

Таким образом, толщина двойного слоя для крупных объектов постоянна и составляет около метра, а для субмикронных – несколько сантиметров и медленно падает с уменьшением их радиуса. Можно ожидать поэтому, что в пылевых облаках, где расстояния между пылинками гораздо меньше, использованное приближение для плотности фотоэлектронов неприменимо.

### 3. Время образования двойного слоя и его толщина

С точностью до обозначений (7) - это та же формула (2), что была использована в работе [Stubbs 2006]. Однако, теперь речь идет об уединенной (но «одетой электронной шубой») частице, и естественно, возникает вопрос о длительности процесса формирования ее заряда. В простейшем приближении мы можем оценить время $\delta t$, в течение которого заряд на освещенной со всех сторон пылинке достигнет значения $q_d^{(0)}$, используя то же значение фототока с поверхности Луны $j_{ph}$=4 10$^{-5}$ A m$^{-2}$:

$$\delta t = \frac{q_d^{(0)}}{4\pi r_d^2 j_{ph}} = \frac{\varepsilon_0}{r_d j_{ph}} V \approx \frac{2.5\cdot 10^{-7}}{r_d} V \text{ s}, \qquad (18)$$

где радиус должен быть выражен в метрах, а потенциал $V \equiv \varphi_0$ – в вольтах. Таким образом, при энергии фотоэлектронов (6) порядка 10 эВ время «зарядки» частицы с радиусом 1 мкм составит около 10 секунд, а с радиусом 10 нм – около 20 минут.

Столь длительное время зарядки требуется из-за огромной поверхностной плотности заряда пылинки. Легко видеть, что для шара с потенциалом $\varphi_0$ всегда выполняется условие

$$\sigma_d = \varepsilon_0 \varphi_0 / r_d, \qquad (19)$$

так что при постоянном потенциале поверхностная плотность заряда $\sigma_d$ обратно пропорциональна радиусу. Очень большой оказывается и напряжённость поля у поверхности пылинки

$$E_d = \sigma_d/\varepsilon_0 = \varphi_0/r_d. \qquad (20)$$

Она составляет около $10^7$ В м$^{-1}$ при радиусе пылинки в 1 мкм и $10^9$ В м$^{-1}$ – при радиусе 10 нм. В то же время для шара большого радиуса (например, для Луны) напряжённость оказывается исчезающе малой.

Казалось бы, возникает парадокс: над заряженной поверхностью Луны электрическое поле практически отсутствует. Однако, этот парадокс легко разрешается. Формулы (19), (20) учитывают только заряд, накапливающийся на поверхности за то время, пока поле слишком мало и не способно удержать выбитые фотоэлектроны, которые улетают навсегда. Плотность возникшего за это время поверхностного заряда действительно пренебрежимо мала. Однако, фотоэлектроны продолжают вылетать с поверхности и после того, как она достигла потенциала $\varphi_0$, только теперь они падают на неё обратно через время $T$ (см. (9), (15)). Возникает двойной электрический слой, и на поверхности при этом появляется дополнительный заряд с плотностью

$$\sigma_S = |q_e|/4\pi r_d^2 = j_{ph}T. \qquad (21)$$

Напряжённость поля, создаваемого этим зарядом и летающими над поверхностью фотоэлектронами, имеет уже значительную величину. Снова полагая, что все фотоэлектроны сконцентрированы в тонком слое на высоте $H_\infty$, см. (15), (16), получаем

$$H_\infty \equiv H(r_d \to \infty) = \sqrt{\frac{\varepsilon_0}{j_{ph}}}\left(\frac{|e|}{2m_e}V^3\right)^{1/4} \approx 0.25 \cdot V^{3/4} \text{ m},$$

$$T_\infty = 4\frac{H_\infty}{v_{ph}} \approx 2\cdot 10^{-6} \cdot V^{1/4} \text{ s}. \qquad (22)$$

Поскольку в этом приближении поле внутри двойного слоя однородно, электрон движется вверх равнозамедленно с начальной скоростью $v_{ph}$. При этом снова предполагается, что фотоэлектроны вылетают только перпендикулярно поверхности. Отметим, что время образования двойного слоя над поверхностью Луны или астероида составляет всего несколько микросекунд, что на много порядков меньше времени зарядки субмикронной пылинки.

### 4. Величина заряда на пылинке

При той же, что у [Stubs2006] величине потенциала $V \approx 4$ В мы получаем из (22) примерно такие же, как в этой работе толщину двойного слоя (около 0.7 м) и напряженность поля внутри слоя (около 6 В/м). Поэтому формула (1) дает примерно ту же величину высоты подъема пылинки

$$h = \frac{3\varepsilon_0}{\rho g}\left(\frac{V}{r_d}\right)^2 \approx 5\cdot 10^{-15}\left(\frac{V}{r}\right)^2, \qquad (23)$$

при подстановке заряда (7), плотности грунта $\rho \approx 3000$ кг м$^{-3}$ и величины ускорения свободного падения на Луне $g \approx 1{,}6$ м с$^{-2}$ ($h$ и $r_d$ выражены в метрах, а $V$ – в вольтах). Это совершенно естественно, поскольку мы используем те же формулы и почти те же значения параметров, что и авторы работы [Stubbs 2006]. Однако, здесь встает принципиально важный вопрос: при каких условиях мы можем подставлять в (1) величину заряда пылинки, определяемую формулой (7)?

Все зависит от того, считаем ли мы, что взлетающие пылинки первоначально лежали на поверхности, и заряд на них появился только после восхода Солнца. Действительно, при большом радиусе $r_d$ (планеты и астероиды) время $\delta t$ (18) практически равно нулю. Следовательно, спустя время $T_\infty$ (22) после начала освещения поверхности поле над ней достигнет величины $\sigma_S / \varepsilon_0$, и после этого уже ни один фотоэлектрон не сможет подняться на высоту, превышающую $H_\infty$. Из (22) видно, что практически для любых возможных значений энергии фотоэлектрона время образования двойного слоя над плоской поверхностью составляет единицы микросекунд, и по истечении этого промежутка заряд на поверхности перестает расти.

В течение этого же времени стабилизируется заряд пылинки, лежащей на поверхности, и оно на несколько порядков меньше времени (18), в течение которого происходит заряд уединенной субмикронной пылинки. Соответственно, заряд $q_{ds} = \pi r_d^2 \sigma_S$, скопившийся на пылинке, лежащей на поверхности, на несколько порядков меньше заряда $q_d^{(0)}$. Поскольку работа выхода фотоэлектрона для пылинок и поверхности одинакова (одно вещество), мы можем записать

$$\frac{\delta t}{T_\infty} = \frac{q_d^{(0)}}{q_{ds}} = \frac{\varepsilon_0 V^{3/2}}{2 r_d j_{ph} H_\infty}\sqrt{\frac{|e|}{2m_e}} \approx \frac{0.1}{r_d}V^{3/4}, \qquad (24)$$

где опять радиус пылинки выражен в метрах, а разность потенциалов – в вольтах.

Следовательно, dynamic fountain mechanism, предложенный в [Stubbs 2006], может функционировать только если пылинки с зарядом порядка $q_d^{(0)}$ (7) к моменту восхода Солнца уже находятся на поверхности (или над ней). Вопрос о том, откуда могли бы взяться эти заряженные пылинки, обсуждается в следующем разделе.

Еще одно проблема возникает в связи со сделанным в [Stubbs 2006] предположении о смене знака и одновременным резким ростом модуля потенциала двойного слоя вблизи терминатора. Именно ростом $|V|$, а с ним и заряда (теперь отрицательного!) пылинок объясняется в [Stubbs 2006] резкое возрастание высоты подъема horizon glow $Z_{max}$ вблизи терминатора. Однако, в этом случае на пылинках должны были бы накапливаться избыточные электроны, и нам не вполне понятно, какие взаимодействия должны удерживать их при напряженности поля у поверхности пылинки в миллионы или даже миллиарды вольт на метр.

## 5. Откуда берутся большие положительные заряды на пылинках?

На вопрос о том, откуда в принципе могли бы взяться сильно заряженные пылинки, ответить совсем не сложно, если учесть, что каждый из постоянно падающих на поверхность метеоритов должен поднимать тучи пыли. На солнечной стороне эта пыль за время полета заряжается, а падая вниз на освещенную поверхность, снова подбрасывается вверх за счет фонтанного механизма. В результате возникает движение вверх-вниз с ростом заряда по дороге.

В течение всего дня такие пылинки непрерывно взлетают и падают, так что имеют достаточно времени, чтобы приобрести соответствующий их размеру заряд порядка $q_d^{(0)}$ (7). На закате мгновенно исчезает двойной электрический слой в местах возникновения теней, и по мере появления все большего количества затененных участков поверхности заряженные пылинки должны оседать на них. Вероятно, прежде всего, заряженная пыль должна оседать в ямах, кратерах, на неосвещенных склонах скал и т.д. В такой ситуации если при ударе о поверхность пылинка под действием электростатических сил не прилипает к ней сразу, то ее отскок вниз, по направлению ко дну кратера или подножию скалы, представляется наиболее вероятным. Если так, то после захода Солнца заряженные пылинки должны быть рассеяны по поверхности крайне неравномерно, образуя более или менее плотные скопления в отдельных местах.

При освещении такого участка на восходе практически мгновенно возникает двойной электрический слой, и вся положительно заряженная пыль должна подбрасываться вверх одновременно. Если эти предпосылки хотя бы частично соответствуют реальной картине, то на восходе пыль не должна подниматься вверх сплошным однородным облаком. Разной мощности фонтаны и фонтанчики пыли должны по мере подъема Солнца над горизонтом взлетать над ямками, провалами, трещинами и кратерами, дна которых достигают солнечные лучи. Напротив, на закате плотность облака пыли, которое днем висело над поверхностью в динамическом равновесии, должна уменьшаться постепенно и более или менее равномерно. Такая картина резко противоречит модели [Stubbs 2006], в которой сила horizon glow на рассвете и на закате должна быть одинаковой.

Однако, для того, чтобы взлетать столбами на рассвете, эта пыль должна сохранить свой заряд в течение лунной ночи, когда, возможно, поток электронов солнечного ветра заряжает поверхность до (отрицательного) потенциала в десятки вольт. Эта оценка основана на значении $1.4 \cdot 10^5$ К [?] эффективной температуры электронов солнечного ветра, что соответствует их кинетической энергии чуть более 10 эВ. Скопления положительно заряженной пыли на поверхности должны были бы притягивать эти электроны, и в течение сотен часов, пока продолжается лунная ночь, пыль должна была бы разрядиться. Однако, решение вопроса о времени жизни положительно заряженной пыли на ночной стороне планеты или астероида представляется достаточно сложной проблемой. Ниже перечислен ряд факторов, которые необходимо учесть при такой оценке.

Прежде всего, подчеркнем, что если бы скорость каждой частицы в солнечном ветре равнялась средней скорости ветра $v_{sw} \approx 4 \cdot 10^5$ м с$^{-1}$, эти частицы вообще не могли бы падать на ночную поверхность. Отклонение направления скорости отдельной частицы от направления $\mathbf{v}_{sw}$ связано с флуктуациями ее поперечной составляющей $\mathbf{v}_\perp$, перпендикулярной направлению на Солнце. Разброс значений именно этой величины (а точнее, связанной с ней кинетической энергии) характеризуется эффективной температурой. Эта температура, грубо говоря, примерно одинакова для электронов и протонов и составляет по порядку величины $10^5$ К (10 эВ) [?]. Такая величина соответствует скорости порядка $2 \cdot 10^6$ м с$^{-1}$ для электронов и $5 \cdot 10^4$ м с$^{-1}$ для протонов. Это означает, что (i) протоны могут попадать только на небольшую часть ночной поверхности вблизи терминатора (см. рис. 1); (ii) при одинаковом направлении протонный ток, текущий к поверхности, пренебрежимо мал по сравнению с электронным.

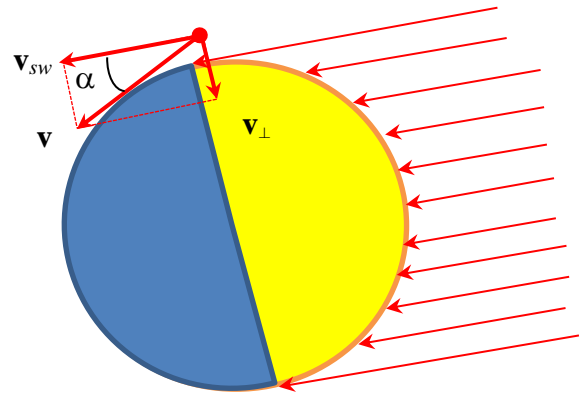

**Рисунок 1.** Частицы, летящие со скоростью $\mathbf{v} = \mathbf{v}_{sw} + \mathbf{v}_\perp$, могут достичь поверхности планеты только в поясе между терминатором и точкой касания $\mathbf{v}$ с поверхностью. Чем меньше угол $\alpha = \mathrm{atan}\left(v_\perp/v_{sw}\right)$, тем уже этот слой. Для протонов солнечного ветра средняя величина этого угла примерно 7°.

Из сказанного следует, что практически только электроны солнечного ветра летят к ночной стороне планеты, и именно из этих электронов состоит зарядовое облако частиц, висящее над поверхностью. Это образование не может быть названо двойным слоем, поскольку и заряд облака, и заряд на поверхности имеют одинаковый (отрицательный) знак. Снова можно предполагать, что основная часть заряда облака сосредоточена в тонком слое, висящем над поверхностью в той области, где скорость летящих к ней электронов падает до нуля. Как плотность заряда этого слоя, так и его высота над поверхностью должны зависеть от того, насколько близко данный район находится к терминатору, и задача определения этих параметров представляет самостоятельный интерес. Тем не менее, даже не решая ее можно утверждать, что для остановки всех электронов солнечного ветра достаточно, чтобы суммарная плотность отрицательного заряда вблизи поверхности достигла примерно величины

$$\sigma = -\frac{\varepsilon_0 V_{sw}}{R} \approx -6\cdot 10^{-16}\ \mathrm{C\ m^{-2}}. \tag{25}$$

Здесь использованы значения $R \approx 1.7\cdot 10^6$ м для радиуса Луны и $|e|V_{sw} \approx 100$ эВ для величины «случайного» вклада в кинетическую энергию электронов солнечного ветра. Плотность (25) пренебрежимо мала по сравнению с плотностью положительного заряда (19) на поверхности субмикронной пылинки $\sigma_d \approx 10^{-11}\varphi_0/r_d \geq 10^{-5}\ \mathrm{C\ m^{-2}}$.

Учитывая также то, что положительно заряженная пыль должна скапливаться в углублениях, куда доступ наклонно падающим электронам затруднен, нельзя исключать возможность того, что ее положительный заряд может сохраняться на протяжении всей ночи.

CONCLUSIONS

Основной полученный выше вывод состоит в том, что пылинки, летающие над освещенной поверхностью Луны, несут на себе чрезвычайно большой положительный заряд. Этот заряд не может возникнуть за счет фотоэффекта на пылинках, лежащих на поверхности, и для того, чтобы пылинки могли взлететь на рассвете, он должен сохраняться даже ночью. Для теоретического решения проблемы «выживания» положительных зарядов на ночной стороне планеты или астероида требуется исследование распределения зарядовой плотности в потоке заряженных частиц, налетающих под разными углами на неоднородно заряженную поверхность.

ЛИТЕРАТУРА